\documentclass[11pt]{amsart}
\usepackage{amssymb}
\usepackage{amscd}
\numberwithin{equation}{section}
\newtheorem{claim}{}[section]

\newtheorem{theorem}[claim]{Theorem}

\newtheorem{corollary}[claim]{Corollary}

\begin{document}

\dedicatory{Dedicated to Gert Pedersen, whose roguish and irrepressible wit is missed by all.}

\title[Matrix Convexity]{New perspectives and Some Celebrated Quantum Inequalities}
\author{Edward G. Effros}
\thanks{Supported by the National Science Foundation DMS-0100883}
\address{Department of Mathematics\\UCLA, Los Angeles, CA 90095-1555}
\email[Edward G. Effros]{ege@math.ucla.edu}

\date{January 28, 2008}
\maketitle
\begin{abstract} Some of the important inequalities associated with quantum entropy are immediate algebraic consequences of the Hansen-Peder\-sen-Jensen inequality. A general argument is given in terms of the matrix perspective of an operator convex function. A matrix analogue of Mar\'{e}chal's extended perspectives provides additional inequalities, including a $p+q\leq 1$ result of Lieb.
\end{abstract}

\section{Introduction}

Several elegant proofs of inequalities due to Lieb  \cite{Lie} and to Lieb and Ruskai \cite{LiR}, have recently appeared (see Nielsen and Petz \cite{Nie}, Ruskai \cite{Rus}). We prove that one can use the ``fully quantized'' Jensen inequality of Frank Hansen and Gert Pedersen \cite{Ha} to eliminate all vestiges of analysis from their bivariable arguments. We then show that a matrix version of Mar\'{e}chal's extended perspectives can be used to formulate more elaborate joint matrix inequalities. In the concluding section we suggest some natural links between matrix convexity theory and the foundations of quantum information theory.

Since the basic difficulties are already apparent in finite
dimensions, we have restricted our attention to finite matrices, and we have avoided any attempt at full
generality even in that context.

I am very much indebted to Mary Beth Ruskai, who corrected a number of errors in my first manuscript, and who made me aware of Lieb's result in the third section.

\section{The classical and matrix notions of perspectives}

Given a convex function $f$ defined on a convex set $K\subseteq \Bbb{R}^{n}$, the \emph{perspective} $g$ is defined on the subset 
\begin{equation*}
L=\left\{ (x,t):t>0\text{ and }x/t\in K\right\} 
\end{equation*}
by 
\begin{equation*}
g(x,t)=f(x/t)t
\end{equation*} 
(see \cite{Hu}). It is a simple exercise to verify that $g(x,t)$ is a jointly  convex function in the sense that if $0\leq c \leq 1$, then
\[
g(cx_{1}+(1-c)x_{2},ct_{1}+(1-c)t_{2})
\leq cg(x_{1},t_{1})+(1-c)g(x_{2},t_{2}). 
\]
An elementary but important example is provided by the continuous convex function
$f(x)=x\log x,$ with $f(0)=0$ defined on $[0,\infty)\subseteq \Bbb{R}$. It follows that the perspective function 
\[
g(x,t)=t\frac{x}{t}\log \frac{x}{t}=x\log x-x\log t
\]
is jointly convex. Letting $p=(p_{i})$ and $q=(q_{i})$ be finite
probability measures with $p_{i}>0$ and $q_{i}>0,$ the convexity of $f$ implies that the classical entropy
\begin{equation*}
H(p)=-\sum p_{i}\log p_{i}
\end{equation*}
is concave, and the convexity of $g$ implies that the relative entropy
\begin{equation*}
(q,p)\mapsto H(q||p)=\sum p_{i}\log p_{i}-p_{i}\log q_{i}
\end{equation*}
is jointly convex on pairs of probability measures.

We recall that if $f:[a,b]\rightarrow \Bbb{R}$ is continuous, and $T$ is an $n\times n$ self-adjoint matrix with spectrum in $[a,b]$, then we can define $%
f_{n}(T)$ by spectral theory (or by using a basis in which $T$ is diagonal). $f$ is said
to be \emph{matrix convex} if for each $n\in \Bbb{N},$ the corresponding
function $f_{n}$ is convex on the self-adjoint $n\times n$ matrices with
spectrum in $[a,b]$. We usually omit the subscript $n$. 

\begin{theorem}[Hansen and Pedersen \cite{Ha}] If $f$ is matrix convex, and $A$ and $B$ are $m\times n$
matrices with $A^{*}A+B^{*}B=I_{n},$ then 
\begin{equation}
f_{n}(A^{*}TA+B^{*}TB)\leq A^{*}f_{m}(T)A+B^{*}f_{m}(T)B.
\end{equation}
\end{theorem}
\noindent We note that their proof does not entail any analysis, but rather is based on a shrewd sequence of matrix manipulations.  As pointed out by Winkler \cite{Wi}, the result may be restated that a real function $f$ on an interval in $\Bbb{R}$ is a matrix convex function if and only if the supergraphs of the $f_{n}$ form a matrix convex system.

We begin with some matrix conventions. Given matrices $L$ and $R$,
we let $[L,R]=LR-RL$. Let us suppose that $L>0$ and $R>0$. If $[L,R]=0$, i.e., the matrices commute, then we may find a basis in which both matrices are diagonalized. It follows that $LR>0$, $[L,R^{-1}]=0$, and we may unambiguously write $\frac{L}{R}$ for the quotient. We also recall that for any continuous function $f,$ $%
f(L)$ commutes with any operator commuting with $L$ (including $L$ itself). Using simultaneously diagonalized matrices, it is evident that we have relations such as $\log LR^{-1} = \log L - \log R$.

\begin{theorem} Suppose that $f$ is operator convex. When restricted to positve commuting matrices, the ``perspective function'' 
\begin{equation}
(L,R)\mapsto g(L,R)=f\left( \frac{L}{R}\right) R
\end{equation}
is jointly convex in the sense that if $[L_{j},R_{j}]=0$ ($j=1,2$),  $L=cL_{1}+(1-c)L_{2}$, $R=cR_{1}+(1-c)R_{2},$ and $0 \leq c \leq 1$, then 
\begin{equation}
g(L,R)\leq cg(L_{1},R_{1})+(1-c)g(L_{2},R_{2}).
\end{equation}
\end{theorem}

\proof The matrices $A=(cR_{1})^{1/2}R^{-1/2}$ and $%
B=((1-c)R_{2})^{1/2}R^{-1/2}$ satisfy $A^{*}A+B^{*}B=I.$ From Theorem 2.1, 
\begin{eqnarray*}
\lefteqn{g(L,R)}\\
&=& Rf\left(\frac{L}{R}\right)\\
&=&R^{1/2}f(R^{-1/2}LR^{-1/2})R^{1/2} \\
&=&R^{1/2}f\left( A^{*}\left( \frac{L_{1}}{R_{1}}\right) A+B^{*}\left( \frac{%
L_{2}}{R_{2}}\right) B\right) R^{1/2} \\
&\leq &R^{1/2}\left( A^{*}f\left( \frac{L_{1}}{R_{1}}\right) A+B^{*}f\left( 
\frac{L_{2}}{R_{2}}\right) B\right) R^{1/2} \\
&=&(cR_{1})^{1/2}f\left( \frac{L_{1}}{R_{1}}\right)
(cR_{1})^{1/2}+((1-c)R_{2})^{1/2}f\left( \frac{L_{2}}{R_{2}}\right)
((1-c)R_{2})^{1/2} \\
&=&cg(L_{1},R_{1})+(1-c)g(L_{2},R_{2}).
\end{eqnarray*}
\endproof

The following is due to Lieb and Ruskai \cite{LiR} (a related early discussion may be found in Lindblad \cite{Lin}).

\begin{corollary}The relative entropy function
\begin{equation*}
(\rho,\sigma )\mapsto S(\rho ||\sigma )=\mathrm{Trace}\,\rho \log \rho -\rho \log \sigma 
\end{equation*}
is jointly convex on the strictly positive $n\times n$ density matrices $%
\rho ,\sigma $.
\end{corollary}

\proof We let $M_{n}$ have the usual Hilbert space structure determined by $%
\langle X,Y\rangle =\mathrm{Trace}$ $XY^{*}.$ Given positive density
matrices $\sigma $ and $\rho ,$ we define operators $R$ and $L$ on $M_{n}$
by $L(X)=\sigma X$ and $R(X)=X\rho .$ Then we have that $L$ and $R$ are
commuting positive operators on the Hilbert space $M_{n}.$ On the other hand
the function $f(x)=x\log x$ is operator convex (see \cite{Bha}, p. 123), and thus
\begin{equation*}
S(\rho ||\sigma )=\langle L\frac{R}{L}\log \frac{R}{L}(I),I\rangle =\langle
g(L,R)(I),I\rangle 
\end{equation*}
is jointly convex. 
\endproof
The following is due to Lieb \cite{Lie}. It was subsequently used by Lieb and Ruskai to prove strong subadditivity for relative entropy \cite{LiR}.

\begin{corollary} If $0<s<1,$ then the function
\begin{equation*}
F(A,B)=\mathrm{Trace}\,A^{s}K^{*}B^{1-s}K
\end{equation*}
is jointly concave on the strictly positive $n\times n$ matrices $A,B$.
\end{corollary}
\proof Since $f(t)=-t^{s}$ is operator convex (see \cite{Bha} Th.5.1.9),  $g(L,R)=-L^{s}R^{1-s}$ is jointly convex for appropriately commuting
operators. Again using the Hilbert space structure on $M_{n},$ we let $L(X)=AX$ and $R(X)=XB.$ It follows that 

\begin{equation*}
(A,B)\mapsto -\mathrm{Trace}\,A^{s}K^{*}B^{1-s}K=\langle g(L,R)(K^{*}),K^{*}\rangle 
\end{equation*}
is jointly convex.
\endproof

Various generalized entropies may be handled in much the same manner.

\section{Mar\'{e}chal's perspectives}
	
P. Mar\'{e}chal has recently introduced an interesting generalization of perspectivity for convex functions \cite{Ma1}, \cite{Ma2}. This also has a natural matrix version. For this purpose we use Hansen and Pedersen's earlier result \cite{Ha1}.

\begin{theorem}  If $f$ is matrix convex, and $f(0)\leq 0,$ and that $A$ and $B$
are $m\times n$ matrices with $A^{*}A+B^{*}B\leq I_{n},$ then 
\begin{equation*}
f_{n}(A^{*}TA+B^{*}TB)\leq A^{*}f_{m}(T)A+B^{*}f_{m}(T)B.
\end{equation*}
\end{theorem}

Given continuous functions $f$ and $h$, and commuting positive matrices $L$ and $R,$
we define
\begin{equation*}
(f\Delta h)(L,R)=f\left(\frac{L}{h(R)}\right)h(R)
\end{equation*}

\begin{theorem} Suppose that $f$ is matrix convex with $f(0)\leq 0$ and that $h$ is matrix concave with $h>0.$ Then $(L,R)\mapsto(f\Delta h)(L,R)$ is jointly
convex on postive commuting matrices $L,R$ in the sense of (2.3). 
\end{theorem}
\proof Let us suppose that $L=cL_{1}+(1-c)L_{2}$ and $R=cR_{1}+(1-c)R_{2}$
where $[L_{j},R_{j}]=0$. We have that $ch(R_{1})+(1-c)h(R_{2})\leq %
h(R),$ hence 
\begin{eqnarray*}
A &=&c^{1/2}h(R_{1})^{1/2}h(R)^{-1/2} \\
B &=&(1-c)^{1/2}h(R_{2})^{1/2}h(R)^{-1/2}
\end{eqnarray*}
satisfy 
\begin{eqnarray*}
\lefteqn{A^{*}A+B^{*}B}\\
&=&h(R)^{-1/2}ch(R_{1})h(R)^{1/2}+h(R)^{-1/2}(1-c)h(R_{2})h(R)^{-1/2}\\
&\leq& h(R)^{-1/2}h(R)h(R)^{-1/2}I=I.
\end{eqnarray*}
It follows from Theorem 3.1 that  
\begin{eqnarray*}
\lefteqn{(f\Delta h)(L,R)}\\
&=&h(R)^{1/2}f(h(R)^{-1/2}Lh(R)^{-1/2})h(R)^{1/2} \\
&=&h(R)^{1/2}f\left(A^{*}\left( \frac{L_{1}}{h(R_{1})}\right)
A+B^{*}\left( \frac{L_{2}}{h(R_{2})}\right) B\right) h(R)^{1/2} \\
&\leq & h(R)^{1/2}A^{*}f\left( \frac{L_{1}}{h(R_{1})}\right)
Ah(R)^{1/2}+h(R)^{1/2}B^{*}f\left( \frac{L_{2}}{h(R_{2})}\right) Bh(R)^{1/2}
\\
&=&ch(R_{1})^{1/2}f\left( \frac{L_{1}}{h(R_{1})}\right)
h(R_{1})^{1/2}+(1-c)h(R_{2})^{1/2}f\left( \frac{L_{2}}{h(R_{2})}\right)
h(R_{2})^{1/2} \\
&=&c(f\Delta h)(L_{1},R_{1})+(1-c)(f\Delta h)(L_{2},R_{2}).
\end{eqnarray*}
\endproof

To illustrate this construction, we reprove a result of Lieb \cite{Lie}.

\begin{corollary} Suppose that $0<p,q$ and that $p+q\leq 1$. Then the function
\begin{equation*}
(A,B)\mapsto \mathrm{Trace}\,A^{q}X^{*}B^{p}X
\end{equation*}
is jointly concave on the positive $n\times n$ matrices.
\end{corollary}

\proof Since $p+q\leq 1$, $p+q$ is a convex combination of $q$ and $1$, i.e., we may choose $0\leq t\leq1$ with $p+q=(1-t)q+t1$. If we let $q=s$, then 
\[
p=-tq+t=(1-q)t=(1-s)t.
\] 
 Thus it suffices to show that if $0\leq s,t \leq 1$, then
\begin{equation*}
(A,B)\mapsto -\mathrm{Trace}\,A^{s}X^{*}B^{(1-s)t}X
\end{equation*}
is jointly convex. The functions $f(x)=-x^{s}$ and $h(y)=y^{t}$ are operator convex and concave, respectively, and 
\[
(f\Delta h)(L,R)=h(R)f\left(\frac{L}{h(R)}\right)=-R^{t}\frac{L^{s}}{R^{st}}=-L^{s}R^{(1-s)t}.
\]
 If we let 
 $L(X)=AX$ and $R(X)=XB$ for $X\in M_{n},$ then it follows from the above theorem that
\begin{equation*}
(A,B)\mapsto -\mathrm{Trace}\,A^{s}X^{*}B^{(1-s)t}X=\langle (f\Delta h)(L,R)(X^{*}),X^{*}%
\rangle
\end{equation*}
is jointly convex.\endproof

\section{matrix convexity}

Perhaps the most interesting aspect of Mar\'{e}chal's construction is that it behaves well under the Fenchel-Legendre transform, and under iteration. S{\o}ren Winkler formulated an analogue of the Fenchel-Legendre duality for matrix convex functions \cite{Wi}, but the transforms are generally set-valued mappings. Further progress might result if one could reformulate his theory in terms of commuting pairs. It should also be noted that other constructions in classical convexity theory, such as the linear fractional transformations of convex functions (see \cite{B}) might also have matrix generalizations.

Until recently the theory of matrix convexity has suffered from a lack of examples and applications. With the advent of quantum information theory (QIT), this situation has dramatically changed. QIT provides a wealth of remarkable, purely non-classical techniques that might clarify some of the conceptual problems in matrix convexity theory. On the other hand, it seems likely that matrix convexity will provide an appropriate framework for many of the calculations in QIT. A striking illustration of this phenomenon can be found in \cite{Dev}.

\end{document}